\documentclass[aps,prd,showpacs,amsmath,amssymb,nofootinbib,twocolumn]{revtex4-1}
\usepackage{graphicx}

\usepackage{bm}

\def\beq{\begin{equation}}
\def\eeq{\end{equation}}

\def\bal{\begin{align}}

\def\eal{\end{align}}
\begin{document}

\title{Hawking radiation in Non local field theories}

\author{Nirmalya Kajuri}
\email{nkajuri@cmi.ac.in}
\affiliation{Chennai Mathematical Institute, Siruseri
Kelambakkam 603103}
\author{Dawood Kothawala}
\email{dawood@physics.iitm.ac.in}
\affiliation{Department of Physics, Indian Institute of Technology Madras, Chennai 600040}

\bibliographystyle{apsrev4-1}

\begin{abstract}

Locality plays a key role in the derivation of Hawking radiation. Could relaxing the locality criterion modify Hawking radiation? We consider Hawking radiation in a general class of non local quantum field theories. We prove that Hawking radiation is unmodified for these theories. Our result establishes the universality of Hawking radiation in a larger class of theories.

\end{abstract}

\maketitle

\section{Introduction.} One of the most striking predictions of quantum field theory in curved spacetime is Hawking radiation - the prediction that quantum effects near the event horizons of black holes cause them to emit radiation. Since Hawking's initial derivation \cite{Hawking:1974sw} the result has been derived in many different ways and appears to be a robust result of quantum field theory. At the same time this result leads to the notorious black hole information loss paradox (see \cite{Mathur:2009hf} for a review) which has remained intractable despite many years of effort.

Locality is one of the key assumptions that leads to this paradox. It has been argued that giving up locality can modify Hawking's result and resolve the black hole information loss paradox. While the possibility of resolving the paradox using non-locality has been studied in the context of black hole complementarity ( originally in \cite{Susskind:1993if, Susskind:1993mu}. See \cite{Papadodimas:2013jku} for a modern version), surprisingly there has been no rigorous attempt at studying Hawking radiation in explicitly non local theories. In this paper we fill this gap.

Non local field theories have been studied for a long time(An early review is \cite{Efimov:1967pjn}). These theories have field equations which involve infinitely many derivatives. Modern work on the subject has focused on non local theories of graviation\cite{Tomboulis:1997gg,Biswas:2010zk,Moffat:2010bh, Biswas:2011ar,Modesto:2011kw, Modesto:2014lga,Frolov:2015bia,Frolov:2015bta} and cosmological models\cite{Deser:2007jk,Barnaby:2007ve,Barnaby:2008fk,Barnaby:2010kx, Deser:2013uya}. We refer the reader to \cite{Tomboulis:2015gfa} for a comprehensive review and \cite{Buoninfante:2018mre} for an interesting discussion of the subject.

In this paper we will consider Hawking radiation in a general class of non local field theories. We will consider theories that are governed by an equation of the form:

\begin{align}
\label{a}
\Box_g(F(\Box_g))\phi =0
\end{align}

where $F$ is an analytic function that is everywhere non zero and $\Box_g$ denotes the d'Alembartian in a general spacetime with metric $g_{\mu \nu}$. An example of such a theory is  $$\Box_g \left(e^{-l^2\Box_g^{2n}}\right) \phi =0$$

where $l$ is some length scale, typically a UV cut-off or a minimal length. Naively, one would expect that the existence of a minimal length modifies Hawking radiation. Our aim in this paper is to rigorously test this hypothesis.

In\cite{Kajuri:2017jmy} Unruh effect had been studied in these theories of the form \eqref{a}. By determining the  Bogoliubov coefficients, it was shown that Unruh effect remained unmodified in these theories. The same result was obtained using the method of Unruh-DeWitt detector\cite{Modesto:2017ycz} and further confirmed in \cite{Gim:2018rcy}, which also explored Lorentz violating theories. These results confirm the universality of Unruh effect in the above class of non local theories. In this paper we show that Hawking radiation is also unmodified. Our derivation follows the same chain of logic as \cite{Kajuri:2017jmy}.

To prove that Hawking radiation is unmodified for the nonlocal theories described above, we follow the derivation of Hawking radiation by Fredenhagen and Haag \cite{Fredenhagen:1989kr}. This is both the cleanest derivation of Hawking radiation (Hawking's original derivation involved the S-matrix, which does not actually exist in this case. See \cite{Wald:1995yp} for a discussion.)and the closest in spirit to the derivation of Unruh effect using Unruh-DeWitt detectors (See also \cite{Fredenhagen:1986jg}). We prove in this paper that for non-local theories which follow, the Fredenhagen-Haag derivation of Hawking radiation goes through completely. 

The possibility of quantum gravity modifications to the Fredenhagen-Haag derivation had been raised in \cite{Jacobson:1991gr} and clarified in \cite{Hambli:1995pp} in the context of a higher (but finite) derivative theory. Our work is close in spirit to the latter as we ask the same questions, but about non local field theories with infinitely many derivatives. 
 
Our result establishes the universality of Hawking radiation in  the general class of non local field theories of the form \eqref{a}. Thus we show that relaxing the condition of locality does not necessarily resolve the black hole information loss paradox. 

We note that the non-locality we consider here is solely in the field theory. In quantum gravity, one expects a breakdown of locality of space-time itself. It is possible that the black hole information paradox may be resolved by considering the non-local effects induced by fluctuations in space-time. This possibility has not been considered in this paper.

Our paper is organized as follows. In the next section, we will present a brief sketch of the derivation of Hawking effect given by Fredenhagen and Haag. We will highlight the key preconditions for the result. In the third section, we will supply proof that the derivation goes through for fields which follow \eqref{a}.

\section{The Fredenhagen-Haag derivation} This section briefly reviews the derivation by Fredenhagen and Haag. See also \cite{Haag:1992hx} and \cite{Salehi:1993gz} for discussions.

We consider a spherically symmetric collapse. Outside the star the metric is Schwarzschild:

\beq
ds^2 = -\left( 1- \frac{2M}{r} \right) + \left( 1- \frac{2M}{r} \right)^{-1} dr^2 +r^2(d \theta^2 + \sin^2 \theta d\phi^2)
\eeq
We will find the following coordinates convenient for our purpose: 

Here $r_0$ denotes the Schwarzschild radius. The $\tau$ coordianate remains meaningful on the horizon. The radius of the star crosses the Scwharzschild radius at $\tau =0$. For finite $\tau$ the horizon is at $r* \rightarrow \infty$
\beq
r* = r+ r_0 \ln \frac{r}{r_0-1}
\eeq
\beq
\tau = t+r* -r
\eeq
We consider a scalar field which satisfies 
\beq
\label{kg}\Box_g \phi =0
\eeq

where $\Box_g$ is the covariant D'Alambertian in this spacetime. 

Now we consider a detector placed in some spacetime region $\mathbb{O}$ far from the horizon at a very late time. We consider the detector to be centered around a point $r=R, t=T$ such that
$$T \gg R, \, \, R \gg r_0$$

The detector will be represented by the observable $C =Q^{*}Q$ where

\beq
Q = \int\, \phi(x) h(x)\sqrt{|g|}d^4x
\eeq
Where $h(x)$ is a smooth function with support in $\mathbb{O}$.

Now for each $t_0 \in [a,b]$ where
$$ a = inf\{t: (t,r,\theta,\phi) \in supp\,h\} $$
$$b = sup\{t: (t,r,\theta,\phi) \in supp\,h \} $$
Corresponding to a given function $h(x)$ there will be a unique solution to the following Cauchy problem: 
\beq
\label{kg2}\Box_g f_{t_0} (x)=0 
\eeq
\beq
\label{cauchy1}\partial_t f_{t_0}(t_0,r,\theta,\phi) = h(t_0,r,\theta,\phi)
\eeq
\beq
\label{cauchy2}f_{t_0}(t_0,r,\theta,\phi)=0
\eeq
This allows us to write 
\beq
\phi(h) = \int\,dt_0 \int_{\Sigma_{t=t_0}} d\sigma^{\mu}\phi(x)\partial_{\mu}f_{t_0}(x) - f_{t_0}(x)\partial_{\mu}\phi(x)
\eeq
Which simplifies to
\beq
\label{inner}\phi(h)= \int_{\Sigma}d\sigma^{\mu}\phi(x)\partial_{\mu}f(x) - f(x)\partial_{\mu}\phi(x) 
\eeq

where 
\beq
f(x) =\int\,dt_0 f_{t_0}(x)
\eeq

As $\phi$ and $f$ both satisfy the Klein Gordon equation, right hand side of \eqref{inner} is independent of $\Sigma$. Therefore we may choose $\Sigma$ to be the surface $\tau =0$. 
We then have 
\begin{align}
\label{master}\notag \langle C\rangle =& \langle \phi(h_1)\phi(h_2)\rangle \\  =&\int d\sigma^{\mu_1} d\sigma^{\mu_2}\langle \phi(x_1)\phi(x_2)\rangle \overleftrightarrow{\partial_{\mu_1}} \overleftrightarrow{\partial_{\mu_2}}f^1(x_1)f^2(x_2)
\end{align}

This equation relates the response rate as measured by the detector to the Wightman function in regions where $f(x)$ and its derivative have support. The only property of the field $\phi(x)$ that we have used in arriving at \eqref{master} is that the field satisfies \eqref{kg}. 

The next step is to find where the solutions of \eqref{kg2} with Cauchy data \eqref{cauchy1},\eqref{cauchy2} have support. For $\tau \geq 0$ the support will lie outside the horizon and one needs to solve the wave equation in the Schwarzschild metric to calculate it. 
If the detector is considered to be at time $T\rightarrow \infty$, on the surface $\tau=0$ $f$ will have support concentrated near $r*\rightarrow \pm \infty$ \cite{Dimock:1987hi}. That is, the function $f$ will split into two parts - one part $f_{-}$ with support arbitrarily near the horizon and the other $f_{+}$ supported arbitrarily near spatial infinity. The right hand side of \eqref{master} will then have contributions from $f_{-}, f_{+}$ and a cross term. If we assume that the two point function is bounded at infinite spacelike separation, then it can be shown that the contribution of the cross term to \eqref{master} vanishes. The contribution of the $f_{+}$ term is independent of the existence of the Black Hole and can be ignored. 

Thus we have the expression for the response function of the detector:
 
\begin{align}
\label{master2} \notag \langle C\rangle =& \langle \phi(h_1)\phi(h_2)\rangle \\  =&\int d\sigma^{\mu_1} d\sigma^{\mu_2}\langle \phi(x_1)\phi(x_2)\rangle \overleftrightarrow{\partial_{\mu_1}} \overleftrightarrow{\partial_{\mu_2}}f^1_{-}(x_1)f^2_{-}(x_2)
\end{align}
which involves the two point function in an arbitrarily short distance neighborhood of $\tau=0, r=r_0$

It is then straightforward to show that $\langle C\rangle$ will vanish unless 
\beq
\label{one}
\lim_{\lambda\rightarrow 0} \lambda^2 \frac{\partial}{\partial x_1^{\mu}}\frac{\partial}{\partial x_2^{\mu}}\langle \phi(\lambda x_1)\phi(\lambda x_2)\rangle \neq 0
\eeq
Which will hold only if the Wightman function $\langle \phi(\lambda x_1)\phi(\lambda x_2)\rangle$ has a short distance singularity. 

This is the key result that we will need. We note that we arrived at this result using the following properties of the field theory: (i) The field $\phi(x)$ satisfies \eqref{kg} (ii) The two point function is bounded.

If one further assumes that the Wightman function near the black hole has leading order singularity structure given by the Hadamard form, one arrives at the result that the response rate of the detector would be that produced by radiation at temperature $\frac{1}{4\pi r_0}$. 

\section{
Non local field theory and Hawking radiation}
The class of non-local theories we consider are given by \eqref{a}

\begin{align}
\Box_g(F(\Box_g))\phi =0
\end{align}

where $F$ is an analytic function that is everywhere non-zero and $\Box_g$ is the d'Alembartian in the same spacetime considered in the previous section.

What are the solutions to this equation? Here we use a key result from the theory of differential equations with infinite derivatives which states that the number of independent solutions of an infinite order differential equation is equal to the number of poles in its propagator\cite{Barnaby:2007ve}. Now from the fact that $F$ is analytic and everywhere non-zero, it follows that for \eqref{a} and for $\Box_g\phi=0$, the number of solutions to the two equations are the same. 

But solutions to $\Box_g\phi=0$ already satisfy \eqref{a}. It follows then that the only solutions to \eqref{a} are the solutions to $\Box_g\phi=0$. 

The next step is to define an inner product on the space of solutions which is invariant under change of co-ordinates. This is given by the usual Klein Gordon inner product 
$$\langle \Phi_1 | \Phi_2 \rangle =  i\int_{\Sigma}\sqrt{h} \mathrm{d^3}x n^\mu(\Phi_1 ^* \partial_{\mu}\Phi_2 -  \Phi_2 \partial_{\mu}\Phi_1^*)$$

where as usual $\Sigma$ is a space-like hypersurface, $h$ is the induced metric on $\Sigma$ and $n^\mu$ is the forward pointing normal on $\Sigma$. It is easy to check that for fields which are solutions of \eqref{a} this inner product is independent of the choice of $\Sigma$.

One might have thought that for an equation involving infinitely many derivatives, the correct inner product would be an infinite-dimensional generalization of the ordinary Klein Gordon inner product. It turns out that the infinite dimensional generalization reduces to the ordinary Klein Gordon inner product on the space of solutions. This is because as all the solutions satisfy $\Box_g\phi=0$, the higher derivative terms in the inner product can be made to vanish.

From the fact that the solutions for the field equations as well as the inner product are the same, we see that all steps up to \eqref{one} go through. 
This shows that the Hawking radiation for non local theories depends only on the short distance  Wightman function of the theory. Note that this is not the case for higher but finite derivative field theories.

We now need to deduce the Wightman function for the theory\eqref{a}. To do this, we note that the Wightman function satisfies the following equation:

$\Box_g\langle \phi(\lambda x_1)\phi(\lambda x_2)\rangle = 0$

Now from the argument above about the solutions to \eqref{a} it follows immediately that the Wightman functions for the two theories are also identical. Note that this is not true for the time ordered correlation functions of the two theories.

Therefore all the steps of the Fredenhaagen-Haag derivation go through unmodified. 

This proves that Hawking radiation is unmodified in the class of non local field theories given by \eqref{a}.

To conclude, we have shown that Hawking radiation is unmodified in a general class of non local field theories. This extends Hawking's result to a wider class of theories. It shows that non-locality of the form considered above (equivalently minimal length in theories of the type \eqref{a}) does not modify Hawking's result and therefore cannot lead to a solution of the black hole information loss paradox.

In the future it will be interesting to check if the result goes through when interacting non local field theories are considered. Whether quantum gravity modifications to locality can resolve the paradox also remains an open question.

\begin{acknowledgements} We would like to thank Leonardo Modesto and L. Sriramkumar for valuable discussions. N.K would also like to thank Alok Laddha and Gaurav Narain for helpful discussions.
\end{acknowledgements} 

\bibliography{hawkingnl}
\end{document}